# A standalone version of IsoFinder for the computational prediction of isochores in genome sequences

Pedro Bernaola-Galván[1], Pedro Carpena[1] and José L. Oliver[2,*]

[1] Dpto. de Física Aplicada II, Universidad de Málaga, and [2] Dpto. de Genética, Facultad de Ciencias, Universidad de Granada, Spain

[*]Corresponding autor

**Abstract**

Isochores are long genome segments relatively homogeneous in G+C. A heuristic algorithm based on entropic segmentation has been developed by our group, and a web server implementing all the required components is available. However, a researcher may want to perform batch processing of many sequences simultaneously in its local machine, instead of analyzing them on one by one basis through the web. To this end, standalone versions are required. We report here the implementation of two standalone programs, able to predict isochores at the sequence level: 1) a command-line version (IsoFinder) for Windows and Linux systems; and 2) a user-friendly version (IsoFinderWin) running under Windows.

Availability: Free download from the IsoFinder project site: http://isofinder.sourceforge.net/
Web server: http://bioinfo2.ugr.es/IsoFinder/
Contact: oliver@ugr.es

**Introduction**

The isochore theory, proposed nearly three decades ago (Macaya, Thiery and Bernardi 1976), accounts for the long-range compositional structures of warm-blooded vertebrates (Bernardi, Olofsson, Filipski, Zerial, Salinas, Cuny, Meunier-Rotival and Rodier 1985). Isochores have been associated to a plethora of genetic and genomic features: gene and retrotransposon densities, staining of chromosome bands, methylation patterns, gene/intron lengths, or tissue specificity of gene products (Bernardi 2004). Given the absence of univocal signals at isochore boundaries, only compositional sequence features can be used for computational isochore prediction.

The relationship between the isochores defined by ultracentrifugation and the genome regions computationally predicted at the sequence level has been explored in many papers, both from our group (Bernaola-Galván, Román-Roldán and Oliver 1996; Li, Bernaola-Galvan, Carpena and Oliver 2003; Li, Bernaola-Galvan, Haghighi and Grosse 2002; Oliver, Bernaola-Galvan, Carpena and Roman-Roldan 2001; Oliver, Carpena, Hackenberg and Bernaola-Galvan 2004; Oliver, Carpena, Roman-Roldan, Mata-Balaguer, Mejias-Romero, Hackenberg and Bernaola-Galvan 2002), as well as from other groups (Costantini, Clay, Auletta and Bernardi 2006; Gao and Zhang 2006; Haiminen and Mannila 2007; Nekrutenko and Li 2000; Zhang, Gao and Zhang 2005). All these papers have shown that many of the biological properties of isochores (gene density, SINE and LINE densities, recombination rate or single nucleotide polymorphism variability) are shared by the computationally predicted, isochore-like regions. Computational methods for isochore finding involving only properties of bulk DNA and no annotated features, third-codon positions, and so on, can uncover regions often fulfilling the properties initially established by ultracentrifugation experiments (Costantini, Clay, Auletta and Bernardi 2006).

The algorithm developed by our group uses entropic segmentation to reliably find isochores in genome sequences; updated descriptions are provided in (Oliver, Carpena, Hackenberg and Bernaola-Galvan 2004) and at the IsoFinder project site (Oliver 2008). The main innovation respect to previous versions of the algorithm (Bernaola-Galván, Román-Roldán and Oliver 1996; Oliver, Bernaola-Galvan, Carpena and Roman-Roldan 2001; Oliver, Carpena, Roman-Roldan, Mata-Balaguer, Mejias-Romero, Hackenberg and Bernaola-Galvan 2002) is the implementation of a coarse-graining technique, enabling to filter out short-range sequence heterogeneity that could mask otherwise the large-scale structure. Recently, the reliability of this algorithm in predicting isochore-like regions at the sequence level has been established through the analysis of large-scale genome patchiness by an independent method: the analysis of the deviations in the power-law behavior of long-range correlations (Oliver, Bernaola-Galvan, Hackenberg and Carpena 2008).

**The algorithm**

To find isochore boundaries, we move a sliding pointer from left to right along the DNA sequence. At each position of the pointer, we compute the mean G+C values to the left and to the right of the pointer. We then determine the position of the pointer for which the difference between left and right mean values (as measured by the *t*-statistic) reaches its maximum. Next, we determine the statistical significance of this potential cutting point, after filtering out short-scale heterogeneities below a given minimum length by applying a coarse-graining technique. Finally, the program checks whether this significance exceeds a probability threshold. If so, the sequence is cut at this point into two subsequences; otherwise, the sequence remains undivided. The procedure continues recursively for each of the two resulting subsequences created by each cut. This leads to the decomposition of a chromosome sequence into long homogeneous genome regions (LHGRs) with well-defined mean G+C contents, each significantly different from the G+C contents of the adjacent LHGRs. Most of these LHGRs can be identified with Bernardi's isochores, given their correlation with biological features such as gene density, SINE and LINE (short, long interspersed repetitive elements) densities, recombination rate or single nucleotide polymorphism variability (Oliver, Bernaola-Galvan, Carpena and Roman-Roldan 2001; Oliver, Carpena, Roman-Roldan, Mata-Balaguer, Mejias-Romero, Hackenberg and Bernaola-Galvan 2002).

## Implementation

Both command-line versions were written in Fortran'90 and compiled using the Lahey/Fujitsu Fortran 95 compiler under Windows XP and Debian Linux. Windows version has been compiled using the Compaq Visual Fortran V.6 compiler under Windows XP.

## Description

Sequences in standard sequence formats (EMBL, GenBank or FASTA) are accepted as input and the program properly manages the islands of N's often present in long genome sequences. The length of the input sequence is only limited by the available memory of the local machine.

The Windows version allows specifying input/output files as well as all the program parameters in the command line, thus facilitating external program calls or its integration into other scripts. Besides the input file, the user should specify a window size for the coarse-graining filter, a significance level to check GC differences between adjacent isochores, and a method to compute the statistical significance. Detailed instructions are provided at http://isofinder.sourceforge.net/. The Windows version (IsoFinderWin) uses the same implementation of the segmentation algorithm as the command line versions and, in addition, provides a user-friendly interface (Fig. 1). It includes interactive graphics which allow the user to inspect the GC profiles of the input sequence at different zoom levels prior to the analysis, superimpose the isochore map on top of the GC profile (Fig. 2), or to obtain information about an isochore by double clicking on the graph. The table listings with isochore coordinates are also interactive windows which allow the user to locate in the graph the desired isochore by double clicking on the list. The exporting capabilities include graphs in BMP format and subsequences from isochores in EMBL, GenBank or FASTA formats.

IsoFinderWin has been developed as a MDI (multiple-document interface) program in order to permit the inspection of several GC profiles, isochore maps or isochore coordinate lists at the same time (Fig. 3).

Both standalone versions of IsoFinder overcome one of the main shortcomings of the web server i.e. the need for uploading the sequence to be analyzed. This process may take a long time when dealing with full chromosome sequences.

Pre-computed isochore maps for many genomes are available at our web site (http://bioinfo2.ugr.es/isochores/), and also at the UCSC Genome Browser (http://genome.cse.ucsc.edu/).

## Acknowledgments

The IsoFinder project was supported by the Spanish Government (Grant No. BIO2005-09116-C03-01) and the Spanish Junta de Andalucía (Grants No. P06-FQM-01858, P07-FQM3163 and CVI-162).

## References

Bernaola-Galván, P., R. Román-Roldán, and J.L. Oliver. 1996. Compositional segmentation and long-range fractal correlations in DNA sequences. *Physical Review E* **53:** 5181-5189.

Bernardi, G. 2004. *Structural and evolutionary genomics. Natural selection in genome evolution*. Elsevier, Amsterdam.

Bernardi, G., B. Olofsson, J. Filipski, M. Zerial, J. Salinas, G. Cuny, M. Meunier-Rotival, and F. Rodier. 1985. The mosaic genome of warm-blooded vertebrates. *Science* **228:** 953-958.

Costantini, M., O. Clay, F. Auletta, and G. Bernardi. 2006. An isochore map of human chromosomes. *Genome Res* **16:** 536-541.

Gao, F. and C.T. Zhang. 2006. GC-Profile: a web-based tool for visualizing and analyzing the variation of GC content in genomic sequences. *Nucleic Acids Res* **34:** W686-691.

Haiminen, N. and H. Mannila. 2007. Discovering isochores by least-squares optimal segmentation. *Gene* **394:** 53-60.

Li, W., P. Bernaola-Galvan, P. Carpena, and J.L. Oliver. 2003. Isochores merit the prefix 'iso'. *Computational Biology and Chemistry* **27:** 5-10.

Li, W., P. Bernaola-Galvan, F. Haghighi, and I. Grosse. 2002. Applications of recursive segmentation to the analysis of DNA sequences. *Comput Chem* **26:** 491-510.

Macaya, G., J.P. Thiery, and G. Bernardi. 1976. An approach to the organization of eukaryotic genomes at a macromolecular level. *J Mol Biol* **108:** 237-254.

Nekrutenko, A. and W.H. Li. 2000. Assessment of compositional heterogeneity within and between eukaryotic genomes. *Genome Res* **10:** 1986-1995.

Oliver, J.L. 2008. IsoFinder Project, pp. IsoFinder: computational prediction of isochores in genome sequences.

Oliver, J.L., P. Bernaola-Galvan, P. Carpena, and R. Roman-Roldan. 2001. Isochore chromosome maps of eukaryotic genomes. *Gene* **276:** 47-56.

Oliver, J.L., P. Bernaola-Galvan, M. Hackenberg, and P. Carpena. 2008. Phylogenetic distribution of large-scale genome patchiness. *BMC Evol Biol* **8:** 107.

Oliver, J.L., P. Carpena, M. Hackenberg, and P. Bernaola-Galvan. 2004. IsoFinder: computational prediction of isochores in genome sequences. *Nucleic Acids Research* **32:** W287-W292.

Oliver, J.L., P. Carpena, R. Roman-Roldan, T. Mata-Balaguer, A. Mejias-Romero, M. Hackenberg, and P. Bernaola-Galvan. 2002. Isochore chromosome maps of the human genome. *Gene* **300:** 117-127.

Zhang, C.T., F. Gao, and R. Zhang. 2005. Segmentation algorithm for DNA sequences. *Phys Rev E Stat Nonlin Soft Matter Phys* **72:** 041917.

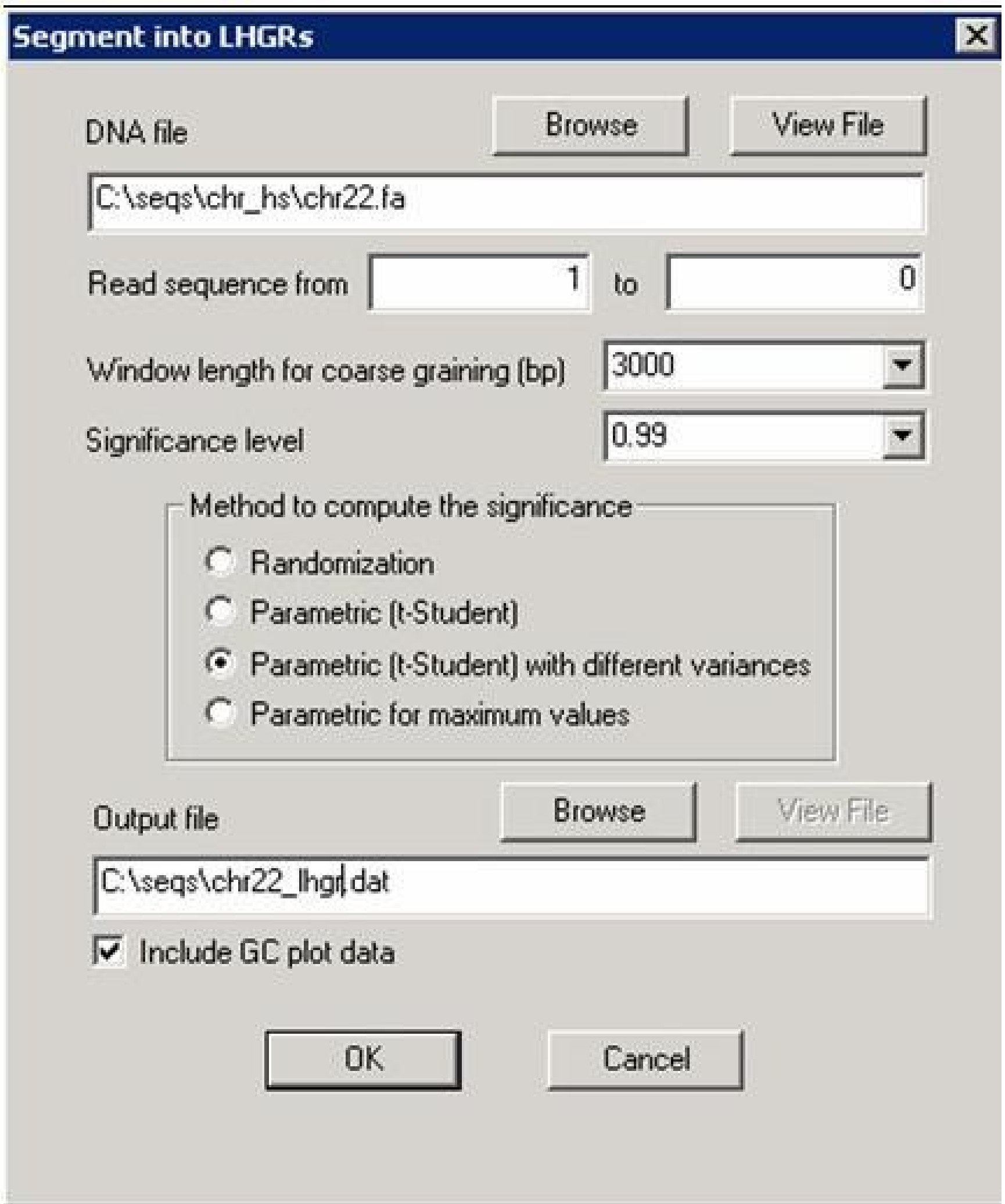

Figure 1. Dialog box for input/output specifications in IsoFinderWin

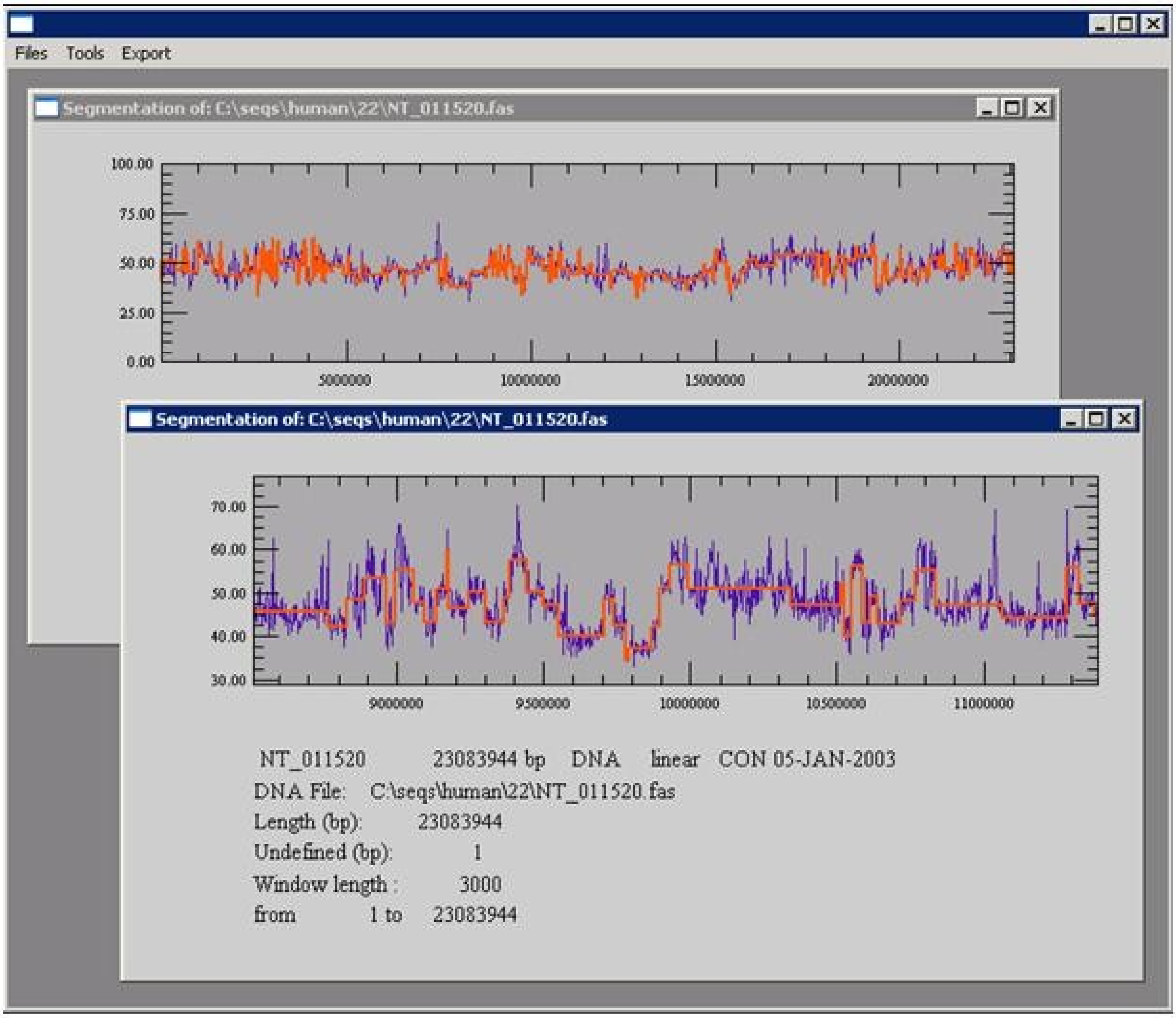

Figure 2. IsoFinderWin multiple-document interface

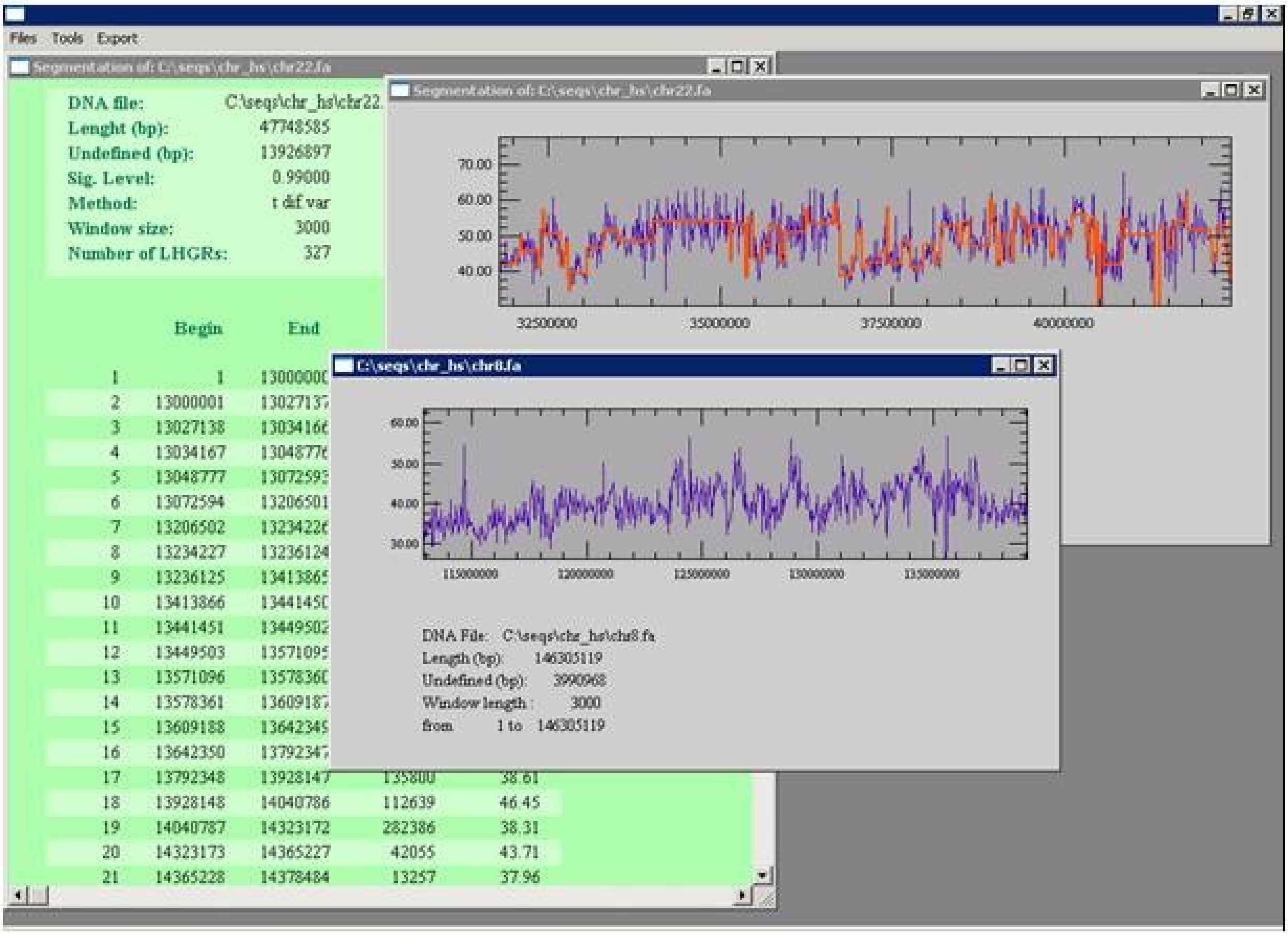

Figure 3. Isochore map of contig NT-011520 (human chromosome 22) at two different zoom scales